\documentclass[
aps,%
12pt,%
final,%
notitlepage,%
oneside,%
onecolumn,%
nobibnotes,%
nofootinbib,%
superscriptaddress,%
noshowpacs,%
centertags]%
{revtex4}

\usepackage{epsfig,subfigure}

\newcommand{\etall}{{\itshape et al.}}
\newcommand{\journal}[4]{{#1}$\;${\bf #2} (#3) #4}
\newcommand{\pr}{\journal {Phys. Rev.}}
\newcommand{\plb}{\journal {Phys. Lett. B}}
\renewcommand{\prd}{\journal {Phys. Rev. D}}
\renewcommand{\prl}{\journal {Phys. Rev. Lett.}}

\newcommand{\npps}{\journal {Nucl. Phys. Proc. Suppl.}}
\newcommand{\nppsb}{\journal {Nucl. Phys. Proc. Suppl. B}}

\newcommand{\prep}{\journal {Phys. Reports}}

\newcommand{\aaj}{\journal {Astron. Astrophys.}}

\renewcommand{\apj}{\journal {Astrophys. J.}}

\newcommand{\arnps}{\journal {Ann. Rev. Nucl. Part. Sci.}}
\newcommand{\nature}{\journal {Nature}}

\newcommand{\nima}{\journal {Nucl. Instr. Meth. A}}
\newcommand{\rpp}{\journal {Rep. Prog. Phys.}}

\newcommand{\app}{\journal {Astropart. Phys.}}
\newcommand{\sjnp}{\journal {Sov. J. Nucl. Phys.}}

\newcommand{\nutau}{\ensuremath{\nu_{\tau}}}
\newcommand{\nue}{\ensuremath{\nu_{{\mathrm e}}}}
\newcommand{\numu}{\ensuremath{\nu_{\mu}}}

\newcommand{\gev}{\ensuremath{{\mathrm{GeV}}}}
\newcommand{\tev}{\ensuremath{{\mathrm{TeV}}}}
\newcommand{\pev}{\ensuremath{{\mathrm{PeV}}}}
\newcommand{\eev}{\ensuremath{{\mathrm{EeV}}}}

\newcommand{\is}{\ensuremath{{\mathrm{s}^{-1}}}}
\newcommand{\isr}{\ensuremath{{\mathrm{sr}^{-1}}}}
\newcommand{\icm}{\ensuremath{{\mathrm{cm}^{-2}}}}
\newcommand{\units}{\gev\,\icm\,\is\,\isr}

\begin{document}
\selectlanguage{english}

\title{From AMANDA to IceCube}

\author{\scriptsize{
M.~Ribordy$^{\dag,m}, $A.~Achterberg$^t$, M.~Ackermann$^d$, J.~Ahrens$^k$, D.W.~Atlee$^h$, J.N.~Bahcall$^{*,u}$, X.~Bai$^a$, B.~Baret$^s$, M.~Bartelt$^n$, R.~Bay$^i$
% ...
\\
(*) Deceased;
(a) Bartol Research Institute, University of Delaware, Newark, DE 19716 USA;
(b) Department of Physics, University of Wuppertal, D-42119 Wuppertal, Germany;
(c) Universit\'e Libre de Bruxelles, Science Faculty CP230, B-1050 Brussels, Belgium;
(d) DESY, D-15735, Zeuthen, Germany;
(e) Blackett Laboratory, Imperial College, London SW7 2BW, UK;
(f) Dept. of Technology, Kalmar University, S-39182 Kalmar, Sweden;
(g) Lawrence Berkeley National Laboratory, Berkeley, CA 94720, USA;
(h) Dept. of Physics, Pennsylvania State University, University Park, PA 16802, USA;
(i) Dept. of Physics, University of California, Berkeley, CA 94720, USA;
(j) Dept. of Physics and Astronomy, University of California, Irvine, CA 92697, USA;
(k) Institute of Physics, University of Mainz, Staudinger Weg 7, D-55099 Mainz, Germany;
(l) Dept. of Physics, University of Maryland, College Park, MD 20742, USA;
(m) University of Mons-Hainaut, 7000 Mons, Belgium;
(n) Dept. of Physics, Universit\"at Dortmund, D-44221 Dortmund, Germany;
(o) Dept. of Physics, University of Wisconsin, Madison, WI 53706, USA;
(p) Dept. of Physics, University of Wisconsin, River Falls, WI 54022, USA;
(q) Division of High Energy Physics, Uppsala University, S-75121 Uppsala, Sweden;
(r) Dept. of Physics, Stockholm University, SE-10691 Stockholm, Sweden;
(s) Vrije Universiteit Brussel, Dienst ELEM, B-1050 Brussels, Belgium;
(t) Dept. of Physics and Astronomy, Utrecht University, NL-3584 CC Utrecht, NL;
(u) Institute for Advanced Study, Princeton, NJ 08540, USA;
(v) Dept. of Physics and Astronomy, University of Kansas, Lawrence, KS 66045, USA;
(w) Dept. of Physics, Southern University, Baton Rouge, LA 70813, USA;
(x) Dept. of Astronomy, University of Wisconsin, Madison, WI 53706, USA;
(y) Dept. of Physics, Chiba University, Chiba 263-8522 Japan;
(z) CTSPS, Clark-Atlanta University, Atlanta, GA 30314, USA;
(aa) Dept. of Astronomy and Astrophysics, Pennsylvania State University, University Park, PA 16802, USA;
(ab) Institut f\"ur Physik, Humboldt Universit\"at zu Berlin, D-12489 Berlin, Germany\\
(\dag) corresponding author: mathieu.ribordy@umh.ac.be\hfill~
}
}

\begin{abstract}{\it
The first string of the neoteric high energy neutrino telescope IceCube successfully
began operating in January 2005. It is anticipated that upon completion the new detector
will vastly increase the sensitivity and extend the reach of AMANDA to higher energies. A
discussion of the IceCube's discovery potential for extra-terrestrial neutrinos, together
with the prospects of new physics derived from the ongoing AMANDA research will be the focus
of this paper.

Preliminary results of the first antarctic high energy neutrino telescope 
AMANDA searching in the muon neutrino channel for localized and diffuse 
excess of extra-terrestrial
neutrinos will be reviewed using data collected between
2000 and 2003. Neutrino flux limits obtained with the all-flavor dedicated UHE and cascade
analyses will be described. A first neutrino spectrum above one TeV in agreement with
atmospheric neutrino flux expectations and no extra-terrestrial contribution will be
presented, followed by a discussion of a limit for neutralino CDM candidates annihilating in
the center of the Sun.
}\end{abstract}

\maketitle

\section{Physics goals}
The AMANDA neutrino telescope was primarily designed to search 
for extra-terrestrial high energy (HE) neutrinos in quest of the mysterious 
origins of the HE cosmic rays (CR).
The occurrence of electronic acceleration up to about 100 TeV has been demonstrated 
in SNR, nevertheless, multiwavelength studies have not established that they may 
accelerate nucleons as well.

Conventional astronomy is continually revealing an ever richer sky at TeV energies, powered 
by astronomical objects such as AGN or SNR and has shown that the cataclysmic GRB phenomena 
are quite common in the universe. It is postulated that these objects are hadronic accelerators 
and may therefore be at the origin of the CR's, up to the highest energies. Most likely, the 
interaction of these accelerated particles with the proton environment and the surrounding 
radiation field results in the production of a secondary HE $\nu$ flux.
The discovery of a positive signal through the exploration of the neutrino sky would provide
an unambiguous signature and begin to resolve a 100 year old controversy.

From the astrophysical perspective, neutrinos provide access to the distant HE sky, 
unavailable to photons which interact with the IR background at TeV energies and with the CMBR 
around 1 PeV, reducing their path from the edge of the Milky Way.
Also, charged cosmic rays are astronomical messengers ({\it i.e.} pointing back to their 
sources) improving with a rigidity increase, but which undergo an energy damping above 
$\approx10^{19.5}$ eV/nucleon due to CMBR photopion production over cosmologically short distances (<50 Mpc).

In the past years, AMANDA has lowered the upper limits of an 
extra-terrestrial HE $\nu$ flux, with no positive signal yet. 
This is why the IceCube project was born.
There are strong arguments in favor of building a magnified version of AMANDA-II.
First, there are guaranteed neutrino sources: the extra-galactic CR interacting with the CMBR produce of order of 1 event/yr/km$^2$, the galactic CR interacting with the ISM in the disk produce an observable flux. Furthermore, a CR excess mostly from the Cygnus region and from the galactic center around $10^{18}$ eV hints at in-flight decay of neutrons primaries after a typical path of a few kpc, producing an observable $\bar\nu_\mathrm{e}$-flux~\cite{halzen-gal-ptsrc}. 
Second, H.E.S.S looking at the TeV SNR RX J1713.7-3946 has demonstrated a clear increase of the flux in the directions of molecular clouds~\cite{hess}, suggesting they may be targets for accelerated protons~\cite{hess-halzen}. In this case, we could expect 20 $\nu_\mu$/yr/km$^2$.
Third, if we consider transparent sources (from which neutrons and pions can escape) and the extra-galactic CR energy density (from the observed CR flux above the knee), the neutrino spectrum can be derived~\cite{wb} leading to an observable flux in IceCube but not in AMANDA-II.
Finally, various models for extra-galactic and galactic point-like emitters predict a signal within reach of the IceCube detector~\cite{halzen-tev-blazar,halzen-gal-ptsrc,guetta-plerion}.
The case of HE neutrino physics potential has been described in more details in e.g.~\cite{review-CR}.

The sensitivity of the IceCube detector upon completion is expected to 
improve by about one order of magnitude over current limits within three 
years and will be well below 
the Waxman-Bahcall upper bound. This may truly open a new window on the universe.

As a bonus, the HE atmospheric neutrino statistics which will be 
accumulated over the years with IceCube will offer us particle physics 
opportunities within and beyond the SM, this will be discussed in section~\ref{sec:prospects}.

\section{The AMANDA and the IceCube neutrino telescopes}
The AMANDA-II detector consists of an array of 677 photomultipliers, 
arranged on 19 strings, instrumenting a cylindrical volume with an outer 
radius of 100 m, buried in the ice under the geographic South Pole.
The detector has been in operation since 2000.
At the deployment depths between 1.5 and 2 km, the ice is 
clear and the secondary muon intensity originating in cosmic air showers in the 
atmosphere is reduced by a large factor. Nevertheless, it still constitutes the
major experimental background, triggering AMANDA at a rate of about 60 Hz.
The 10 inner strings, arranged on a cylinder with a 60 m radius, is called AMANDA-B10 
and was completed in 1997~\cite{nature}. The proof of principle was demonstrated 
in observing atmospheric neutrinos~\cite{amanda97_atmnu} at the expected levels.

Events are reconstructed~\cite{amanda-reco} by measuring the arrival time of \v{C}erenkov light propagating in the transparent medium, emitted by either crossing relativistic (neutrino-induced) muons or by showers resulting from neutral current interactions for all three neutrino flavors, \nue, \numu~and \nutau~and from \nue~and \nutau~charged current interaction (referred to as the cascade channels in the following).
AMANDA is therefore sensitive to all three neutrino flavors, which is important in the context of neutrino oscillations,
as we expect all three flavors to be populated equally \nue:\numu:\nutau=1:1:1 after traveling cosmological distance from their source (\nue:\numu:\nutau=1:2:$\epsilon$, $\epsilon<10^{-5}$ is expected from an hadronic accelerator)~\cite{cosmnumixing}.
The neutrino-induced muon channel and the cascade channels have different specificities and are then complementary: while muon tracks are reconstructed with an incidence direction resolution $\Delta\Psi\!\approx\! 2.5^\circ$, cascade events are reconstructed with a better energy resolution $\Delta\mathrm{log}E\!\approx\! 15\%$.

Analyses with AMANDA are facing various sources of systematic errors. 
They originate in uncertainties concerning the ice properties, the absolute detector 
sensitivities and the primary CR ray spectrum normalization and its exact composition.
Specifically, UHE AMANDA analyses must also account for the uncertainties in neutrino 
cross section and muon propagation.

IceCube's first string was successfully deployed at a maximal depth of 2.45 km 
under the South Pole ice cap in January 2005.
The observatory will eventually consist of an array of 80 strings encompassing the existing AMANDA-II detector.
The one km long strings will be equipped with 60 digital optical modules each,
arranged in the knots of a triangular lattice with 125 m edge in a one km$^3$ 
hexagonal volume. This is more than two orders of magnitude larger than AMANDA-II
and prospect analyses results~\cite{henrike} show that the sensitivities will be rescaled by
roughly the same factor, with a slightly shifted energy threshold though ($E_\nu \approx 100$ GeV), 
due to a sparser instrumentation of the volume. Sensitivities in all detection channels
will extend to higher energies and neutrino signals will be probed up to $E_\nu>10$ EeV. 
Events will be identified in the cascade channels above 10 TeV with the exciting possibility of 
observing double bang $\nu_\tau$-events around one PeV. We will report more precisely on the 
neutrino-induced muon channel later and discuss the discovery potential of IceCube (section~\ref{sec:prospects}).
The trigger rate is expected to be around 1700 Hz, while atmospheric neutrino-induced
muon events will be recorded at the pace of 300 per day, with an improved angular accuracy
$\Delta\Psi\!<\! 1^\circ$ compared to AMANDA-II capabilities.

On top of each IceCube string, at the surface, elements of the IceTop array are being deployed. 
It will eventually consist of 160 ice \v{C}erenkov detectors and will allow the detection of air
showers at energies above $\approx$ 1 PeV. This will not only facilitate the IceCube background rejection, but
will also enable precise composition studies of the charged CR spectrum in the knee region.

\section{Results from AMANDA}\label{sec:results}
We focus in this section on recent results from searches for an extra-terrestrial neutrino flux, from individual point-like emitters
and from a cosmological distribution of unresolved weaker sources.

\subsection{Searches for an extra-terrestrial diffuse neutrino flux}
The underlying theoretical models for diffuse HE $\nu$ flux are often linked to the UHE CR flux~\cite{mpr,wb,ssq}, from which neutrino flux upper bounds are derived providing benchmarks for our searches. These analyses aim at finding global excess, exploiting an expected harder extra-terrestrial neutrino spectrum ($\mathrm{d}\Phi/\mathrm{d}E \!\!\sim\!\! E^{-\alpha}, \alpha\!\!\approx\!\! 2$) in comparison to the steeply falling atmospheric neutrino background ($\alpha\! =\! 3.7$). They critically depend on detector simulation closely matching the experimental data. All detection channels are available to perform these searches, providing an internal cross check of our results.
Below final and preliminary results are presented, 90\% C.L. upper limits 
and sensitivities will be quoted using the all-flavor scheme (the total neutrino flux at Earth in our results assuming \nue:\numu:\nutau=1:1:1, $\nu/\bar\nu=1$) and assuming $E^{-2}$ benchmark neutrino spectral shapes.
The next two results are quickly summarized, as they have been previously discussed in some detail in~\cite{moriond04} and corresponding papers are now available~\cite{uhe97,casc00}. We then exhibit the first muon neutrino spectrum above one TeV and extract from it an upper limit for an extra-terrestrial flux.

\begin{figure}[h]
\setcaptionmargin{5mm}\onelinecaptionstrue

\begin{minipage}{0.45\linewidth}

\includegraphics[width=0.8\textwidth]{plots/uhe97sens.ps2}
\captionstyle{normal}\caption{\scriptsize{MC results of the individual 
neutrino flavor contributions for a UHE diffuse $E^{-2}$ neutrino flux 
at the quoted strength~(eq.~\ref{uhe97limit}).\hfill~}}
\label{fig:uhe97}
\end{minipage}\hfill\begin{minipage}{0.45\linewidth}

\includegraphics[width=0.8\textwidth]{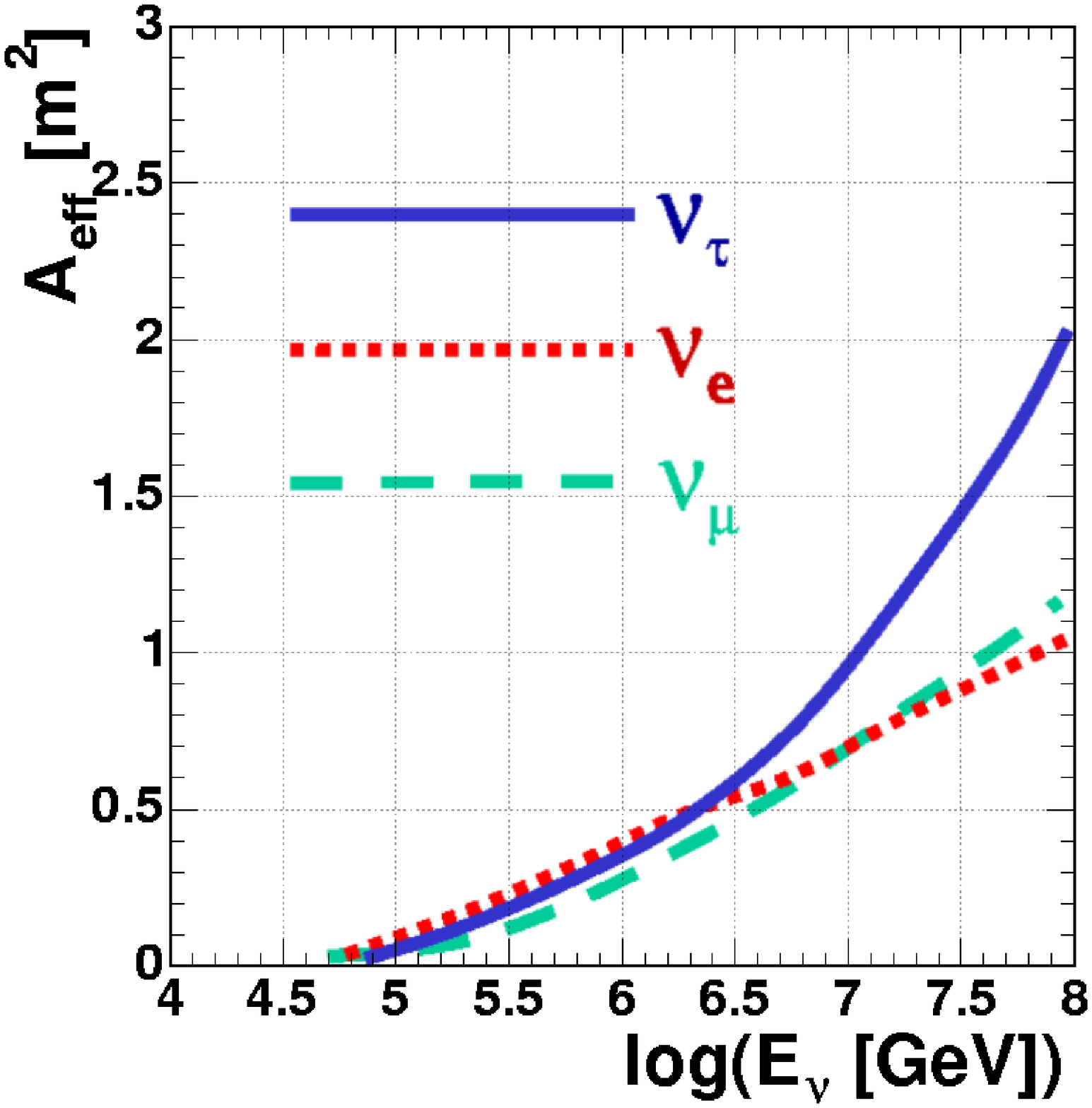}
\captionstyle{normal}\caption{\scriptsize{
The effective area of the 2000 cascade analysis.\hfill~}}
\label{fig:casc00}
\end{minipage}

\end{figure}

A search for an UHE extra-terrestrial neutrino flux using AMANDA-B10 '97 data set yield an all-flavor limit~\cite{uhe97}:
\begin{equation}
\label{uhe97limit}
E_\nu^2 \,{\mathrm{d}\Phi_\nu \over \mathrm{d}E_\nu} = 0.99 \cdot 10^{-6}\,\units, \,\,\,1\,\pev < E_\nu< 3\,\eev.
\end{equation}
In this range of energy, because of the rise of the neutrino cross sections, events concentrate near the horizon
and illuminate the whole detector. Therefore new selection techniques were developed, in order to discriminate
neutrino events from the HE atmospheric background. This analysis has shown similar sensitivity to all 
three neutrino flavors, although slightly higher for muon neutrinos, as shown in Fig.~\ref{fig:uhe97}.

A search in the cascade channels, using the 2000 AMANDA-II 
data set~\cite{casc00} has lead to an all-flavor upper limit:
$$E_\nu^2\, {\mathrm{d}\Phi_\nu \over \mathrm{d}E_\nu} = 0.86 \cdot 10^{-6} \,\units, \,\,\,50\,\tev<E_\nu<5\,\pev,$$
extracted from the effective area plot shown in Fig.~\ref{fig:casc00}.
The analysis demonstrated the all-flavor capabilities of the AMANDA-II detector, showing similar sensitivities in all three
flavors.
Based on the neutrino effective area of a specific analysis, 
the relevance of specific models can be assessed for an arbitrary spectral shape. 
This particular search did exclude various models according to the model rejection factor $\mu_{90\%}/n_\mathrm{model}$~\cite{hill-mrf}.
$\mu_{90\%}$ is the upper limit on the number of signal events, including the background and the systematic uncertainties 
and $n_\mathrm{model}$ is the average number of signal events expected for a differential neutrino flux, 
described by $\mathrm{d} \Phi^{\nu_\alpha}_\mathrm{model}/\mathrm{d}E_{\nu_\alpha}/\mathrm{d}\Omega$.
$n_\mathrm{model}$ is calculated according to
\begin{equation}
n_\mathrm{model}= \sum_{\alpha=\mathrm{e},\mu,\tau} T_\mathrm{life} \int_\Omega\int_{E_{\nu_\alpha}} A^{\nu_\alpha}_\mathrm{eff}(E_{\nu_\alpha},\delta) \frac{\mathrm{d} \Phi^{\nu_\alpha}_\mathrm{model}}{\mathrm{d}E_{\nu_\alpha} \mathrm{d}\Omega}\mathrm{d}E_{\nu_\alpha} \mathrm{d}\Omega,
\label{eq:nsig}
\end{equation}
in which $T_\mathrm{life}$ is the detector lifetime.
The integral over $\mathrm{d}\Omega$ contributes a factor $4\pi$ in this specific analysis because an isotropic diffuse flux is assumed (the dependency of $A^{\nu_\alpha}_\mathrm{eff}(E_{\nu_\alpha},\delta)$ on the declination $\delta$ is irrelevant in this specific diffuse flux search analysis).
Specific models, P~p$\gamma$~\cite{ppgamma} and MPR~\cite{mpr} were not discarded by the results of this analysis. 
However, we show now that the P p$\gamma$ hypothetical neutrino flux model does not survive the preliminary results of a 4-year combined prospective analysis.

The preliminary result of a combined 4-year analysis of data collected between 2000 and 2003, encompassing 807 days of detector lifetime and conducted in the muon neutrino channel sets an all-flavor sensitivity~\cite{diffuse4yr-icrc05}:
$$E_\nu^2 {d\Phi_\nu \over dE_\nu} = 3.3 \cdot 10^{-7} \,\units,\,\,\,16\,\tev<E_\nu<2\,\pev.$$
The number of experimental neutrino-induced muon events which has been observed is 6 with an expectation of 9.7,
yielding an upper flux limit of:
$$E_\nu^2 {d\Phi_\nu \over dE_\nu} = 1.29 \cdot 10^{-7} \,\units,\,\,\,16\,\tev<E_\nu<2\,\pev.$$
These results do not however include sources of systematic errors and should be cautiously appreciated at the moment.
This upper flux limit is approximately half an order of magnitude above the WB upper bound and, if confirmed,
rules out P p$\gamma$ and the most optimistic MPR diffuse neutrino flux models.

The preliminary result from an UHE analysis with AMANDA-II, from data collected in 2000, sets an all-flavor sensitivity~\cite{uhe-icrc05}:
$$E_\nu^2 {d\Phi_\nu \over dE_\nu} = 3.8 \cdot 10^{-7} \,\units,\,\,180\,\tev<E_\nu<1.8\,\eev$$
By combining 5 years of data, the upper flux limit is expected to be further constrained: 
$$E_\nu^2 {d\Phi_\nu \over dE_\nu} = 0.93 \cdot 10^{-7} \,\units,\,\,180\,\tev<E_\nu<1.8\,\eev.$$

\begin{figure}
\setcaptionmargin{5mm}\onelinecaptionstrue

\begin{minipage}{0.4\linewidth}
\includegraphics[width=\textwidth]{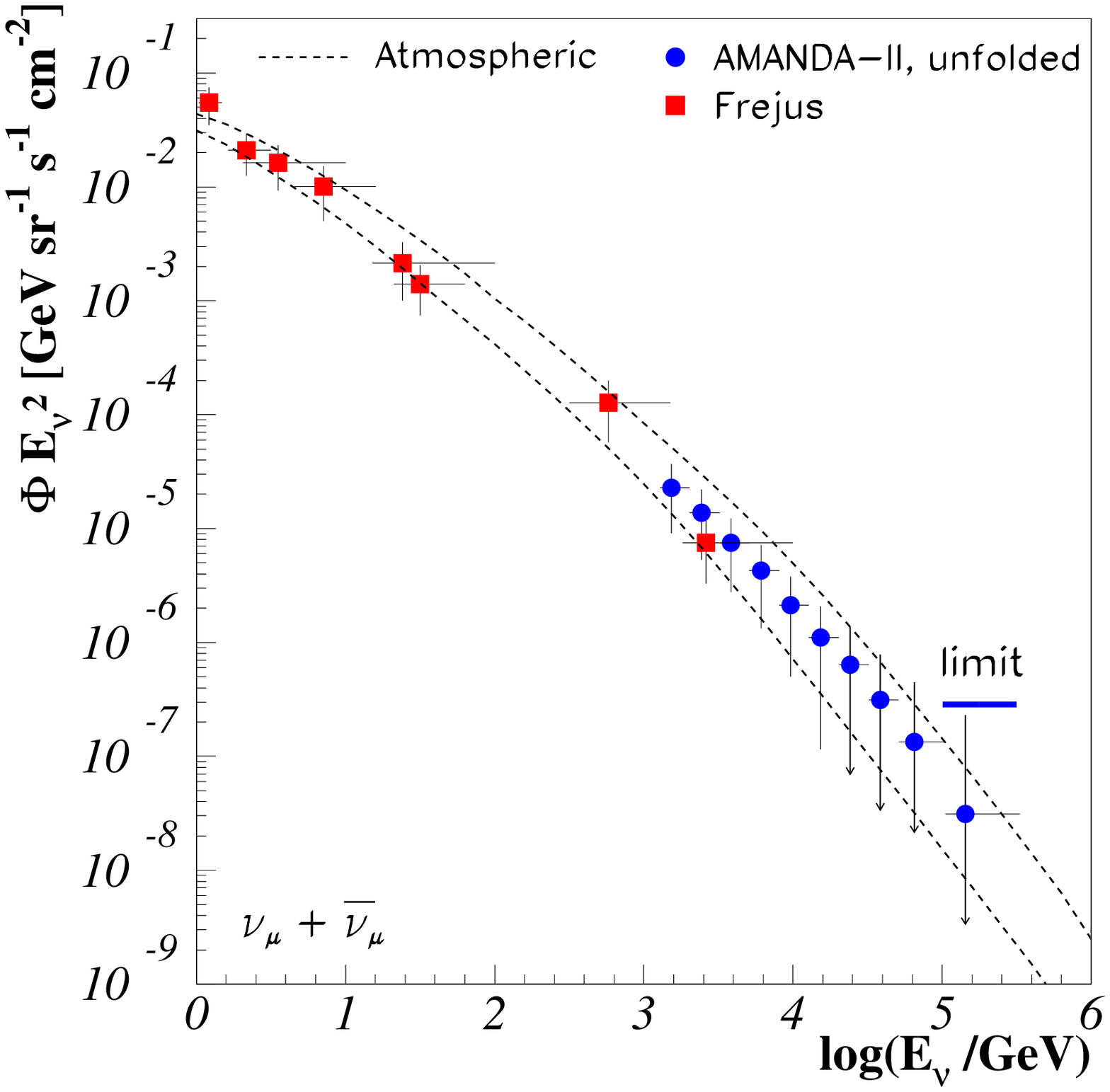}
\captionstyle{normal}\caption{\scriptsize{The AMANDA-II reconstructed HE neutrino spectrum. 
Results from the Frejus experiment~\cite{frejus} are also shown. 
The dotted curves are the predictions for horizontal and vertical fluxes, 
parametrized according to Volkowa above 100 GeV and Honda below 100 GeV~\cite{atmnu-para}.\hfill~}}\label{fig:nuspectrum}

\end{minipage}\hfill\begin{minipage}{0.6\linewidth}

\includegraphics[width=0.8\textwidth]{plots/diffuseflux-summary-plot.ps2}
\captionstyle{normal} \caption{\scriptsize{Diffuse point source flux summary. Lines labeled 1-4 are the upper limits from the '97 $\nu_\mu$ analysis~\cite{diff97}, the '97 UHE analysis~\cite{uhe97}, the '00 cascade analysis~\cite{casc00} and resulting from the '00 atmospheric neutrino spectrum~\cite{nuspectrum-icrc05}. Line 5-6 are the preliminary sensitivities from the '00 UHE analysis~\cite{uhe-icrc05} and the AMANDA-II 4-year combined $\nu_\mu$ analysis~\cite{diffuse4yr-icrc05}. Line 7 shows the AMANDA-II projected sensitivity of a 5-year combined UHE analysis. Also shown the WB~\cite{wb} and MPR~\cite{mpr} upper bounds and the flux predictions of ``S\&S''~\cite{ssq} and ``MPR'' models.}\hfill~}
\label{fig:diffuse-summary}

\end{minipage}

\end{figure}

We will now discuss a different approach in our search for an extra-terrestrial 
diffuse excess. Using the data collected between 2000 and 2003, the 
atmospheric muon neutrino spectrum has been reconstructed above one TeV, 
shown in Fig.~\ref{fig:nuspectrum}. To take the finite 
detector energy resolution into account, the spectrum has been unfolded, using a 
regularization procedure to cope with the low statistics in the HE 
bins. The method used to determine a possible extra-terrestrial contribution is 
the following:
from a set of generated $E^{-2}$ signal spectra, the neutrino event 
distributions (after regularized unfolding) in each individual bin are 
constructed for various flux strengths. Now, we consider various energy ranges.
An upper limit can be calculated, using Feldman-Cousins unified scheme~\cite{fc},
from the distributions of the integrated number of MC signal events and the 
corresponding number of events in the experimental spectrum. 
The best all-flavor limit to an extra-terrestrial contribution that we obtain 
is~\cite{nuspectrum-icrc05}:
$$E^2 {d\Phi_\nu \over dE_\nu} = 7.8 \cdot 10^{-7}\,\units,\,\,\,100\,\tev<E_\nu<300\,\tev.$$
This HE muon neutrino spectrum offers other physics opportunities mentioned in section~\ref{sec:prospects}.

Results of these various diffuse neutrino analyses are summarized in Fig.~\ref{fig:diffuse-summary}.

\subsection{Searches for localized neutrino flux excesses}
This section briefly illustrates analyses searching for a statistical excess originating in
narrow regions of the northern sky. These analyses are exclusively conducted in the muon channel, because of its better pointing resolution. The sensitivities of these analyses are optimized by taking advantage of the experimentally observed off-source detector response which defines the background.

\subsubsection{Searches for steady point sources in the northern sky}\label{sec:ptsrc}
\begin{figure}
\setcaptionmargin{5mm}\onelinecaptionstrue
\includegraphics[width=0.45\textwidth]{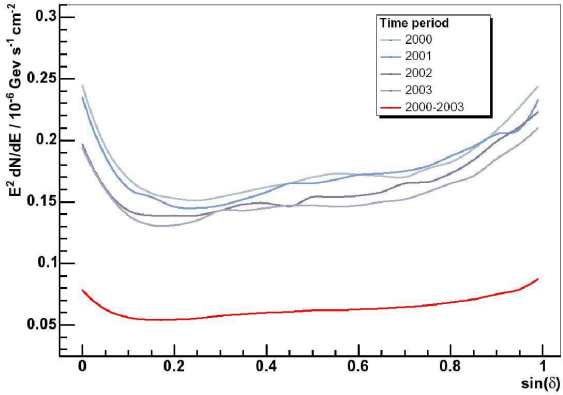}
\hspace{0.02\textwidth}
\includegraphics[width=0.45\textwidth]{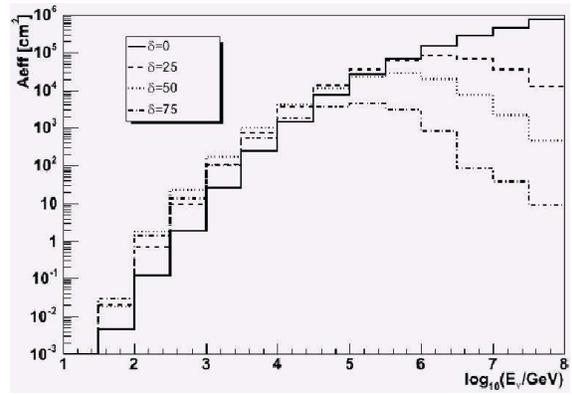}
\captionstyle{normal} \caption{\scriptsize{Preliminary results of the 4-year combined point source search. 
Left: sensitivity.
Right: muon neutrino effective area (for various $\delta$).}\hfill~
}
\label{fig:ptsrc-0003}
\end{figure}

The sensitivity to point sources has been increased by a large factor in the recent years, using the growing set of accumulated AMANDA data, profiting from consequent developments in the analyses methods and an improved understanding of the detector. Previous analyses~\cite{pt97,pt00,pt00_02} demonstrated the course of these progresses. 
The preliminary results~\cite{steady-icrc05}
from a combined 4-year point source search show
a rather uniform sensitivity w.r.t. the declination $\delta$, 
also remarkably close to the horizon (Fig.~\ref{fig:ptsrc-0003}, left).
The declination averaged sensitivity is
$$
E_{\nu_\mu}^2 {\mathrm{d}\Phi_{\nu_\mu} \over \mathrm{d}E_{\nu_\mu}} \approx 0.7\cdot 10^{-7}\,\gev\,\icm\,\is,\,\,90\%\,\mathrm{C.L.}.
$$
The muon neutrino effective area $A^\nu_\mathrm{eff}$ shown on the right of Fig.~\ref{fig:ptsrc-0003} allows us to calculate the limit for an arbitrary flux spectrum according to eq.~\ref{eq:nsig} (with $A^{\nu_\alpha}_\mathrm{eff}(E_{\nu_\alpha},\delta) =0, \,\alpha=\mathrm{e},\tau$).
The final neutrino sample consists of 3329 up-going neutrino-induced muon tracks, at an estimated 90\% purity level. Two approaches have been considered in order to study the event clustering: a search in the whole northern sky and a search on 33 preselected point source candidates, believed to be hadronic accelerators. 
This two approaches differ by their associated trial factor, which is taken into account when calculating a possible excess significance.
No statistically significant departures from the null signal hypothesis have been found.
The preselected source sample results are summarized Table~\ref{table}.

\subsubsection{Search for a solar neutralino dark matter annihilation signature}\label{subsection:wimp}
\begin{figure}

\begin{minipage}{0.45\linewidth}

\setcaptionmargin{5mm}\onelinecaptionstrue
\includegraphics[width=0.8\textwidth]{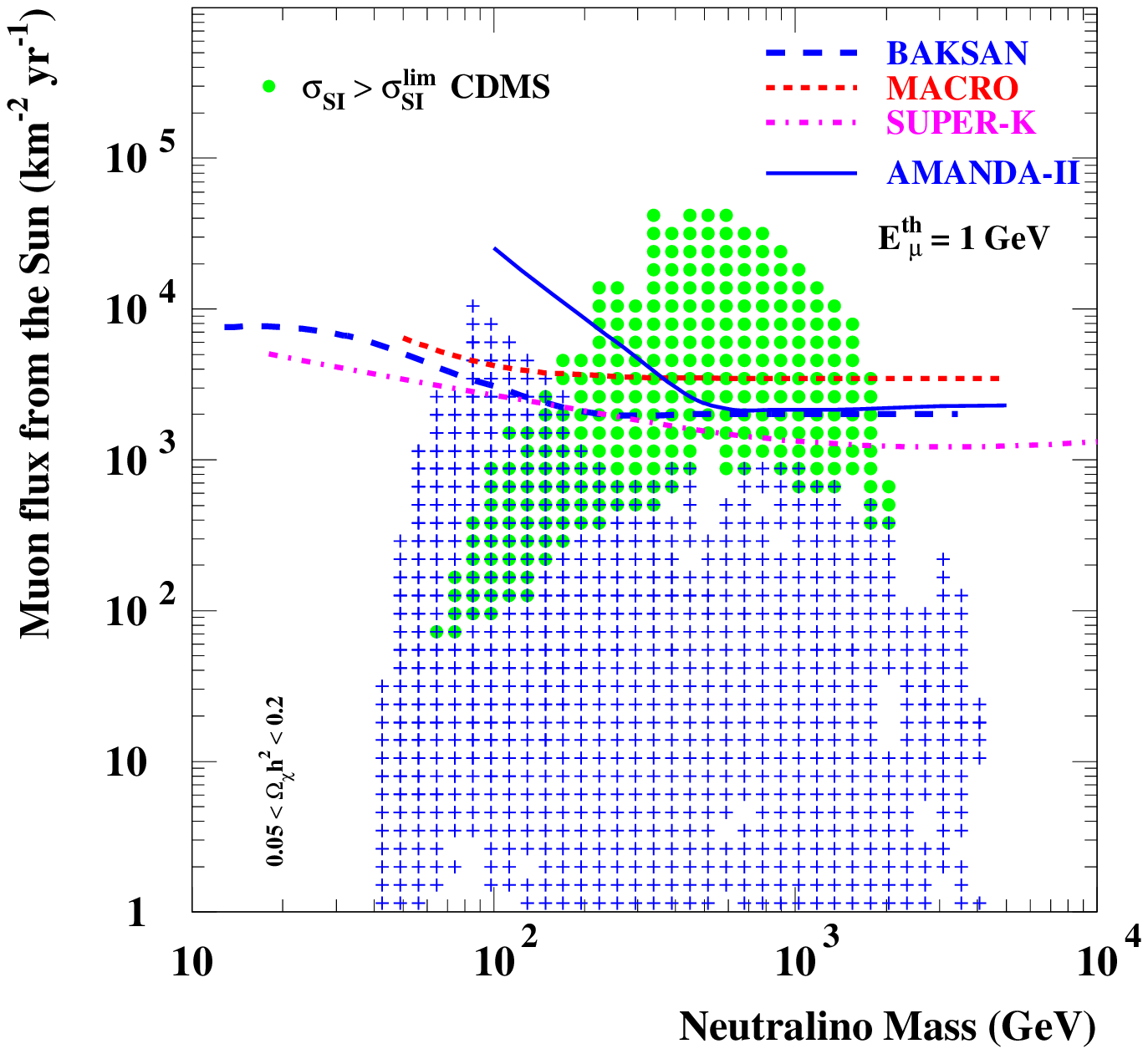}
\captionstyle{normal} \caption{\scriptsize{$\Phi_\mu$ vs $m_\chi$ in AMANDA-II for various MSSM models satisfying $0.05\!\!<\!\!\Omega h^2\!\!<\!\! 0.2$. Dots (crosses) are for models which are (not yet) excluded.}\hfill~}
\label{fig:solar-wimp}

\end{minipage}\hfill\begin{minipage}{0.45\linewidth}

\setcaptionmargin{5mm}\onelinecaptionstrue
\includegraphics[width=0.8\textwidth]{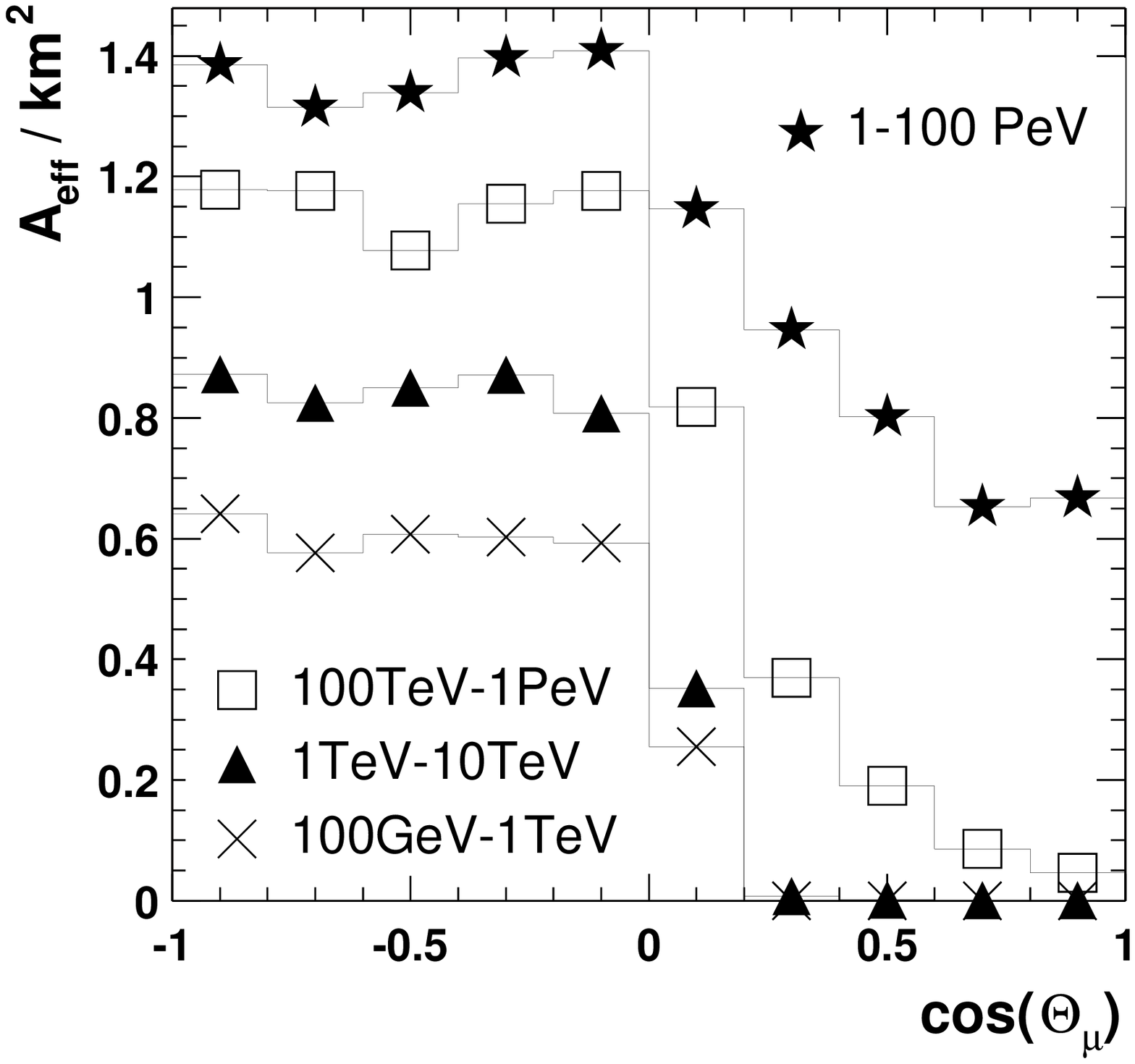}
\captionstyle{normal} \caption{\scriptsize{IceCube neutrino-induced muon effective area w.r.t. the declination, for various energy ranges.}\hfill~}
\label{fig:i3-effarea}

\end{minipage}

\end{figure}

One of the most urgent cosmological problem comes from the non observation of 
$\approx$80\% of the matter content of the universe. 
This DM question can be explored with HE neutrino telescopes, in the case
of non baryonic CDM in the form of the lightest neutralino in the MSSM
framework:
neutralinos in 
the halo scattering off ordinary matter may eventually be gravitationally 
trapped in macroscopic objects. Subsequent scatterings then reduce their 
velocity resulting in their accumulation over astronomical times in the core 
of these objects, where they annihilate pairwise producing neutrinos that can 
be detected.
AMANDA-B10 has performed indirect searches for non baryonic CDM, from the center of the Earth~\cite{wimp}.
The improved reconstruction capabilities of AMANDA-II for nearly horizontal tracks allows a similar search for CDM from the Sun.
During 143.7 days of the detector lifetime 2001, the Sun was located below the horizon. A solar neutralino WIMP analysis has recently been completed~\cite{solar-wimp}. The exclusion limit, shown in Fig.~\ref{fig:solar-wimp}, is slightly above the level of direct CDM search experiments~\cite{cdms-result}. 
Once a few years of data has been cumulated, the WIMP sensitivity is expected to explore the MSSM parameter space close to the reach of the current direct CDM search experiment.
Nevertheless, it must be stressed that nuclear recoil and indirect CDM search experiments are not equivalent.
In the case of an unevenly distributed CDM halo throughout the galaxy,
the latter have a detection potential which remains intact, which renders these indirect searches interesting.
Moreover, solar spin dependent scattering cross sections could significantly enhance the trapping rate of neutralinos.
It is worth mentioning that IceCube with a gain of about one order of 
magnitude in sensitivity will be able to cover a large unexplored region of the MSSM parameter space.

\subsection{Other physics analyses}
\noindent Not all analysis topics in progress have been covered here. Notably:

- The CR composition, studied using data from AMANDA-B10 and including SPASE-2 information, is discussed in~\cite{amanda_spase_cr}.

- A stacking AGN point source analysis will be completed shortly, which consists of a selection of various source samples, belonging to different phenomenological classes, characterized by their morphology, their luminosity and their spectrum.
Preliminary results, using the 4-year combined point source sample, did not show any significant
deviations over background expectations~\cite{stacking-icrc05}.

- A search for transient point sources is in progress. It consists in a selection of potential candidates for hadronic acceleration, based on the exceptional variability of their multiwavelength spectra, as observed with conventional astronomy. This enhancement could be correlated with periods of enhanced neutrino emission.
Preliminary results~\cite{transient-icrc05} did not reveal any excess so far. 

\section{Physics prospects with IceCube}\label{sec:prospects}
IceCube will reach unmatched sensitivities in searching for extra-terrestrial
HE neutrinos, according to the detailed simulation\footnote{
Using the AMANDA-II MC. However, the hardware performance of IceCube is expected to be better: 
background rejection, HE reconstruction and angular resolution will probably be significantly improved.
} results of prospective 
analyses in the neutrino-induced muon channel~\cite{henrike}. 
This is reported in the next two sections.
IceCube will also collect a set of the order of one million neutrinos over 10 years,
in the range 100 GeV$< E_\nu <10$ PeV. This powerful HE 
$\nu$ beam, through the study of its energy and angular distribution,
will allow for instance to calibrate the detector, to determine (or set upper limits to) the HE charmed particle production cross sections~\cite{prompt} and to probe physics beyond the SM, as e.g 
non standard (exotic) neutrino oscillation scenarios.

\subsection{Diffuse extra-terrestrial flux sensitivity}
After 3 years of operation, IceCube is expected to reach an all-flavor sensitivity of
$$E_\nu^2 {\mathrm{d}\Phi_\nu \over \mathrm{d}E_\nu} = 4.2 \cdot 10^{-9}\,\units,\,\,100\,\tev<E_\nu<100\,\pev,\,\,90\%\,\mathrm{C.L.},$$
more than half an order of magnitude below the WB upper bound (see Fig.~\ref{fig:diffuse-summary}). 
This will allow us to confront the predictions from more conservative models of hadronic 
acceleration~\cite{ssq,wb,mpr}\footnote{
Conservatively assuming RQPM~\cite{rqpm} HE atmospheric neutrino background flux from prompt decays, instead
of the standard TIG pQCD~\cite{tig}, in which case our limit would improve by factor two.
}.

\subsection{Point source flux sensitivity}
The all-flavor point source sensitivity after 3 years of data taking, 
will be at the level 
$$
E_{\nu_\mu}^2 {\mathrm{d}\Phi_{\nu_\mu} \over \mathrm{d}E_{\nu_\mu}} = 7.2 \cdot 10^{-9}\,\gev\,\icm\,\is\,\,E>5\,\mathrm{TeV},\,\,90\%\mathrm{C.L.},
$$
This cannot be simply compared to the AMANDA-II 4-year combined analysis, because of the different energy threshold of the analyses. But, if we restrict ourselves to the assumption of a $E^{-2}$ benchmark spectrum with energies above $\approx$1~TeV, this represents a gain in sensitivity by more than one order of magnitude, in reach of various model predictions (see Introduction).
Fig.~\ref{fig:i3-effarea} shows that already at ``low'' energy, 100~GeV$<E<$1~TeV, the effective area reaches 0.6 km$^2$. Above 10 PeV, the Southern sky is opening and the galactic center can be probed with an effective area of as much as 0.5 km$^2$.

\subsection{Physics beyond the SM}
In a world with $\delta+4$ dimensions, the Planck scale can be considerably lowered, down to the TeV-scale, solving the hierarchy problem, this huge existing gap between the Planck and the electroweak scales~\cite{microBH-hierarchy}. Phenomenology predicts in this case the copious production of micro black holes which democratically evaporate into SM particles~\cite{microBH-pheno}. In neutrino telescopes, distinct signatures in energy and angular spectra from the atmospheric ones could be expected. These prospects have been studied in detail in~\cite{microBH_i3}, profiting from the UHE neutrino fluxes, guaranteed cosmogenic~\cite{cosmogenic} or presumed (WB~\cite{wb}).

In the case of a violation of the equivalence principle (non universal coupling of neutrino flavors to gravity) or of the Lorentz invariance (flavor-dependent maximal attainable velocities), the pattern of neutrino oscillations can be modified~\cite{vli-original}: in addition to mass-induced oscillations, other HE oscillations induced by ``new physics'' could be present. 
Consider the oscillation probability, in the case of mass eigenstates coinciding with ``new physics'' eigenstates:
\begin{equation}
\label{eq:vli}
P_{\nu_\mu \rightarrow \nu_\mu}=1-\sin^2{2\theta}\sin^2{\Bigl(\bigl({\Delta m^2 \over 4E}+{\Delta \delta E \over 2}\bigr)L\Bigr)}.
\end{equation}
It is immediately apparent from the second argument of eq.~\ref{eq:vli} that the mass term is strongly suppressed at HE.
Prospects using HE atmospheric neutrinos collected after 10 years of IceCube have been studied in~\cite{vli-IceCube}: clear signatures up to relative violation effects at the $\Delta\delta < 10^{-28}$ level, well beyond the reach of other detectors ($\Delta\delta \lesssim 10^{-26}$~\cite{vli-combined,vli-MACRO} ) are visible by looking at the distortions of the angular and of the energy spectra, as compared to pure mass-induced oscillation expectations.

\section{Conclusions}
AMANDA-II limits to an extra-terrestrial diffuse flux are tightening down close to 
the WB upper bounds leading to strong constraints on hadronic acceleration models. 
Searches for individual point sources of neutrinos have not revealed any excess so far, 
but our continually improving sensitivities may still divulge the first extra-terrestrial neutrinos, 
if so, IceCube would begin an era of precision measurements of the neutrino sky. 
If AMANDA-II does not discover point sources, 
IceCube will nonetheless contribute 
to particle physics within and beyond the SM, 
first extra-terrestrial neutrinos will be observed 
and it is likely, 
based on robust arguments reviewed here,
that first sources will shimmer through IceCube.

\begin{acknowledgments}
\scriptsize{
This research is supported by the following agencies: National Science Foundation-Office of Polar Programs, National Science Foundation-Physics Division, University of Wisconsin Alumni Research Foundation, Department of Energy, and National Energy Research Scientific Computing Center, UC-Irvine AENEAS Supercomputer Facility, USA; Swedish Research Council, Swedish Polar Research Secretariat and Knut and Alice Wallenberg Foundation, Sweden; German Ministry for Education and Research, Deutsche Forschungsgemeinschaft (DFG), Germany; Fund for Scientific Research (FNRS-FWO), Flanders Institute to encourage scientific and technological research in industry (IWT) and Belgian Federal Office for Scientific, Technical and Cultural affairs (OSTC), Belgium. D.F.C. acknowledges the support of the NSF CAREER program. M. Ribordy acknowledges the support of the Swiss National Research Foundation.}
\end{acknowledgments}

\begin{table}[p]
\setcaptionmargin{0mm} \onelinecaptionsfalse
\captionstyle{flushleft} \caption{\scriptsize{Preliminary results from the 4-year combined point source search for a selected sample of sources. $\delta$, $\alpha$, $n_\mathrm{obs}$ and $n_\mathrm{b}$ are respectively the declination, the right ascension, the number of observed events and the expected background.$\Phi_{90\%}$ is the integrated flux upper limit above 10 \gev, assuming a spectral index $\alpha=2$ in $10^{-8}$ cm$^{-2}$ s$^{-1}$ units.}\hfill~}
\bigskip
{\scriptsize
\begin{tabular}{lccccc|lccccc}
\hline\hline
candidate~~~~~~~~ & ~~$\delta$ [$^o$]~~ & ~~$\alpha$ [hr]~~ & ~~$n_\mathrm{obs}$~~ & ~~$n_\mathrm{b}$~~ & ~~$\Phi_{90\%}$~~ & ~candidate~~~~~~~~ & ~~$\delta$ [$^o$]~~ & ~~$\alpha$ [hr]~~ & ~~$n_\mathrm{obs}$~~ & ~~$n_\mathrm{b}$~~ & ~~$\Phi_{90\%}$\\
\hline
\multicolumn{12}{c}{\bf TeV Blazars}\\
Markarian 421 & 38.2 & 11.07 & 6 & 5.6 & 0.68 & ~1ES 2344+514 & 51.7 & 23.78 & 3 & 4.9 & 0.38\\
Markarian 501 & 39.8 & 16.90 & 5 & 5.0 & 0.61 & ~1ES 1959+650 & 65.1 & 20.00 & 5 & 3.7 & 1.0\\
1ES 1426+428 & 42.7 & 14.48 & 4 & 4.3 & 0.54\\
\multicolumn{12}{c}{\bf GeV Blazars}\\
QSO 0528+134 & 13.4 & 5.52 & 4 & 5.0 & 0.39 & ~QSO 0219+428 & 42.9 & 2.38 & 4 & 4.3 & 0.54\\
QSO 0235+164 & 16.6 & 2.62 & 6 & 5.0 & 0.70 & ~QSO 0954+556 & 55.0 & 9.87 & 2 & 5.2 & 0.22\\
QSO 1611+343 & 34.4 & 16.24 & 5 & 5.2 & 0.56 & ~QSO 0716+714 & 71.3 & 7.36 & 1 & 3.3 & 0.30\\
QSO 1633+382 & 38.2 & 16.59 & 4 & 5.6 & 0.37\\
\multicolumn{12}{c}{\bf Microquasars}\\
SS433 & 5.0 & 19.20 & 2 & 4.5 & 0.21 & ~Cygnus X3 & 41.0 & 20.54 & 6 & 5.0 & 0.77\\
GRS 1915+105 & 10.9 & 19.25 & 6 & 4.8 & 0.71 & ~XTE J1118+480 & 48.0 & 11.30 & 2 & 5.4 & 0.20\\
GRO J0422+32 & 32.9 & 4.36 & 5 & 5.1 & 0.59 & ~CI Cam & 56.0 & 4.33 & 5 & 5.1 & 0.66\\
Cygnus X1 & 35.2 & 19.97 & 4 & 5.2 & 0.40 & ~LS I +61 303 & 61.2 & 2.68 & 3 & 3.7 & 0.60\\
\multicolumn{12}{c}{\bf SNR \& Pulsars}\\
SGR 1900+14 & 9.3 & 19.12 & 3 & 4.3 & 0.35 & ~Crab Nebula & 22.0 & 5.58 & 10 & 5.4 & 1.3\\
Geminga & 17.9 & 6.57 & 3 & 5.2 & 0.29 & ~Cassiopeia A & 58.8 & 23.39 & 4 & 4.6 & 0.57\\
\multicolumn{12}{c}{\bf Miscellaneous}\\
3EG J0450+1105 & 11.4 & 4.82 & 6 & 4.7 & 0.72 & ~J2032+4131 & 41.5 & 20.54 & 6 & 5.3 & 0.74\\
M 87 & 12.4 & 12.51 & 4 & 4.9 & 0.39 & ~NGC 1275 & 41.5 & 3.33 & 4 & 5.3 & 0.41\\
UHE CR Doublet & 20.4 & 1.28 & 3 & 5.1 & 0.30 & ~UHE CR Triplet & 56.9 & 11.32 & 6 & 4.7 & 0.95\\
AO 0535+26 & 26.3 & 5.65 & 5 & 5.0 & 0.57 & ~PSR J0205+6449 & 64.8 & 2.09 & 1 & 3.7 & 0.24\\
PSR 1951+32 & 32.9 & 19.88 & 2 & 5.1 & 0.21\\
\hline\hline
\end{tabular}
}
\label{table}
\end{table}

\begin{thebibliography}{99}
\scriptsize{
\bibitem{halzen-gal-ptsrc} L.A.~Anchordoqui, H.~Goldberg, F.~Halzen, T.J.~Weiler, \plb{593}{04}{42}.
\bibitem{hess} H.E.S.S. coll., \nature{432}{04}{75}.
\bibitem{hess-halzen} J.~Alvarez-Muniz, F.~Halzen, \apj{576}{02}{L33}.
\bibitem{wb} E.~Waxman, J.~Bahcall, \prd{59}{99}{023002}; E.~Waxman, J.~Bahcall, \prd{64}{01}{023002}.
\bibitem{halzen-tev-blazar} F.~Halzen, D.~Hooper, \app{23}{05}{537}.
\bibitem{guetta-plerion} D.~Guetta, E.~Amato, \app{19}{03}{403}; E.~Amato, D.~Guetta, P.~Blasi, \aaj{402}{03}{827}.
\bibitem{review-CR} T.K.~Gaisser, F.~Halzen, T.~Stanev, \prep{258}{95}{173}; J.G.~Learned, K.~Mannheim, \arnps{50}{00}{679}; D.F.~Torres, L.A.~Anchordoqui, \rpp{67}{04}{1663}; F.~Halzen, D.~Hooper, \rpp{65}{02}{1025}.
\bibitem{nature} E.~Andr\'es \etall, \nature{410}{01}{441}.
\bibitem{amanda97_atmnu} J. Ahrens et al., \prd{66}{02}{012005}.
\bibitem{amanda-reco} J.~Ahrens \etall, \nima{524}{04}{169}; G.~Japaridze, M.~Ribordy, arXiv:astro-ph/0506136.
\bibitem{cosmnumixing} H.~Athar, M.~Jezabek, O.~Yasuda, \pr{D62}{00}{103007}.
\bibitem{henrike} J.~Ahrens \etall, \app{20}{04}{507}
\bibitem{mpr} K.~Mannheim, R.J.~Protheroe, J.P.~Rachen, \pr{D63}{01}{023003}.
\bibitem{ssq}F.W.~Stecker, M.H.~Salamon, \journal{Space Sci. Rev.}{75}{96}{341}.
\bibitem{moriond04} M. Ribordy et al., proc. of the 39th rencontres de Moriond, La Thuile, 2004, arXiv:hep-ex/0405035.
\bibitem{uhe97} M. Ackermann et al., \app{22}{05}{339}.
\bibitem{casc00} M. Ackermann et al., \app{22}{04}{127}.
\bibitem{hill-mrf} G.~Hill, K.~Rawlins, \app{19}{03}{393}.
\bibitem{ppgamma} R.J.~Protheroe, [astro-ph/9612213].
\bibitem{diffuse4yr-icrc05}J.~Hodges \etall, in proc. of 29th ICRC, Pune, 2005.
\bibitem{uhe-icrc05}L.~Gerhardt \etall, in proc. of 29th ICRC, Pune, 2005.
\bibitem{fc}G.J.~Feldman, R.D.~Cousins, \prd{57}{98}{3873}.
\bibitem{nuspectrum-icrc05}K.~Muenich \etall, in proc. of 29th ICRC, Pune, 2005.
\bibitem{diff97} J. Ahrens \etall, \prl{90}{03}{251101}.
\bibitem{atmnu-para} L.V.~Volkowa, \sjnp{31}{80}{784}; L.V.~Volkova, G.T.~Zatsepin, \sjnp{37}{80}{212}; M.~Honda \etall, \prd{52}{95}{4985}.
\bibitem{frejus} K.~Daum, W.~Rhode, the Frejus Collaboration, \journal{Zeitschrift für Physik C}{66}{95}{177}.

\bibitem{pt00} J. Ahrens \etall, \prl{92}{04}{071102}.
\bibitem{pt97} J. Ahrens \etall, \apj{583}{03}{1040}.
\bibitem{pt00_02} M. Ackermann \etall, \prd{71}{05}{077102}.
\bibitem{steady-icrc05}M.~Ackermann, E.~Bernardini, T.~Hauschildt \etall, in proc. of 29th ICRC, Pune, 2005.
\bibitem{wimp}J. Ahrens \etall, \prd{66}{02}{032006}; M. Ackermann \etall, in preparation.
\bibitem{solar-wimp}M. Ackermann \etall, submitted to Astropart. Phys.
\bibitem{cdms-result}R.W.~Schnee, \npps{143}{2005}{429}.


\bibitem{amanda_spase_cr} J. Ahrens \etall, \app{21}{04}{565}; C.~P\'eres~de~los~Heros \etall, \nppsb{\bf 136}{04}{85}.

\bibitem{stacking-icrc05}A.~Gross, T.~Messarius \etall, in proc. of 29th ICRC, Pune, 2005.
\bibitem{transient-icrc05}M.~Ackermann, E.~Bernardini, T.~Hauschildt, E.~Resconi \etall, in proc. of 29th ICRC, Pune, 2005.

\bibitem{prompt}P. Lipari, \ap{1}{93}{195}; C.G.S.~Costa, F.~Halzen, C.~Salles, \prd{66}{02}{113002};
\bibitem{rqpm}E.V.~Bugaev \etall, \prd{58}{98}{054001}.
\bibitem{tig}M.~Thunman, G.~Ingelman, P.~Gondolo, \app{5}{96}{309}.
\bibitem{microBH-hierarchy}
N.~Arkani-Hamed, S.~Dimopoulos, G.R.~Dvali, \plb{429}{98}{263}; I.~Antoniadis, N.~Arkani-Hamed, S.~Dimopoulos, G.R.~Dvali, \plb{436}{98}{257}; R.~Emparan, G.T.~Horowitz, R.C.~Myers, \prl{85}{00}{499}.
\bibitem{microBH-pheno}
S.B.~Giddings, S.~Thomas, \prd{65}{02}{056010}; S.~Dimopoulos, G.~Landsberg, \prl{87}{01}{161602}.
\bibitem{microBH_i3}
J.~Alvarez-Muniz, J.L.~Feng, F.~Halzen, T.~Han, D.~Hooper, \prd{65}{02}{124015}; M.~Kowalski, A.~Ringwald, H.~Tu, \plb{529}{02}{1}.
\bibitem{cosmogenic}
T.~Stanev, R.~Engel, A.~Mucke, R.J.~Protheroe, J.P.~Rachen, \prd{62}{00}{093005}; R.~Engel, D.~Seckel, T.~Stanev, \prd{64}{01}{093010}; O.E.~Kalashev, V.A.~Kuzmin, D.V.~Semikoz, G.~Sigl, \prd{66}{02}{063004}.


\bibitem{vli-original}M.~Gasperini, \prd{38}{88}{2635}; S.R.~Coleman, S.L.~Glashow, \plb{405}{97}{249}; S.R.~Coleman, S.L.~Glashow, \prd{59}{99}{116008}.
\bibitem{vli-IceCube} 
M.C. GonzalezGarcia, F. Halzen, M. Maltoni, \pr{D71}{05}{093010}. 
\bibitem {vli-combined} 
M.C. Gonzalez-Garcia, M. Maltoni, \pr{D70}{04}{033010}. 
\bibitem{vli-MACRO} 
G. Battistoni et al., \plb{615}{05}{14}. 

}\end{thebibliography}
\end{document}